\documentclass[12pt,rotating]{elsart}
\usepackage{rotating}

\usepackage{cites,setspace,url}
\usepackage{epsfig}

\bibnum{k}

\begin{document}
\citenum{}

\begin{frontmatter}

\title{Remark on the studies of the muon-induced neutron background in the liquid scintillator detectors}

\author{Karim ZBIRI}
\corauth{Tel.: +1 215 895 1887; +1 215 895 2990.\\
E-mail address: (K. Zbiri)zbiri@physics.drexel.edu
}

\maketitle

\address{Department of physics, Drexel University, 3141 Chestnut St,
Philadelphia, PA 19104, USA}

\begin{abstract}
This article gives a point of view on the studies of the muon-induced
background for the underground experiments using a liquid scintillator
detectors. The results obtained are in good agreement with the experimental
 data, especially for the muon-induced neutron yield in the liquid scintillator.


\end{abstract}

\end{frontmatter}

\section{Introduction}

In the underground located experiments, the cosmic muons are one of the main
sources of background, which made their study a crucial step.
 When they are moving through a medium, such as rock or water, these highly energetic particles are initiating
electromagnetic and hadronic showers before reaching the detector, and once they enter the detector they start showering again.    Therefore; the showering process not only in the detector, but also in the environment of the detector is important and has to be taken in account when making simulation as it will be discussed in the following.\\
For several years different Monte Carlo Codes have been developed, and two
of the most used in particle physics studies are FLUKA \cite{Fluka} and Geant4 \cite{G4}.
The models used in these codes were intensively tested and validated before
they have been released for each Monte Carlo code. Various studies
were made to investigate the muon-induced neutron production
in the liquid scintillator \cite{Wang01,Kudry03,Arau05,Mei06}, but the simulations showed a discrepancy of at least $30\%$ when compared to the data. This article
presents a point of view on doing the Monte Carlo studies, aiming to reconcile the simulations with the measurements.
The Monte Carlo code used in this study is Geant4 (release 9.0 patch 02), the
 physics processes  selected for the simulations are similar to these described in Ref.
 \cite{Arau05} except for the gamma interactions where a standard electromagnetic
package is used instead of ``Low Energy'' one.

\section{Muon-induced neutron production}

\begin{figure}
\includegraphics[scale=0.7]{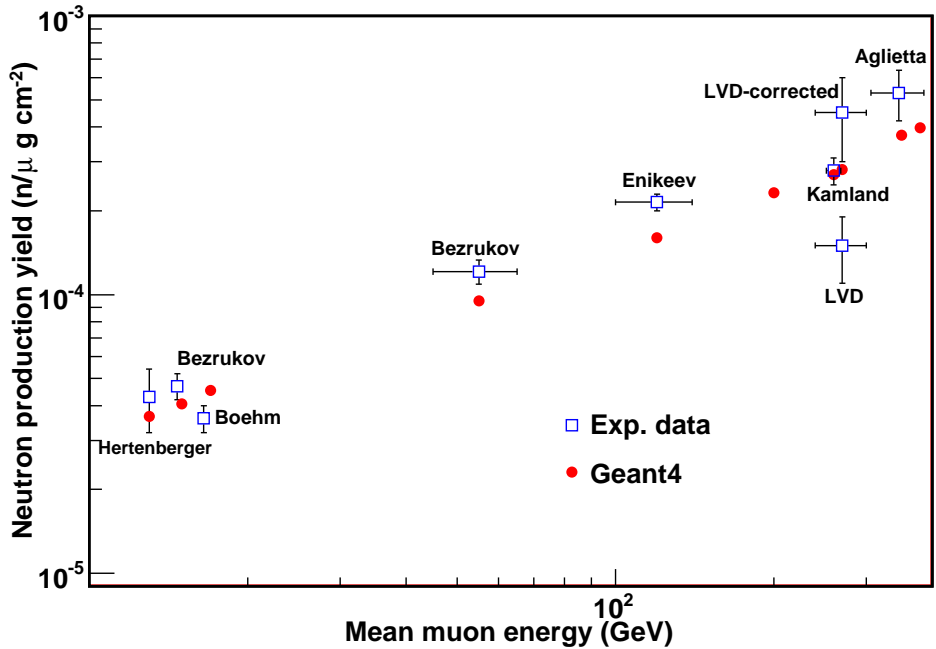}
\caption{Neutron Yield as function of muon energy for $C_{12}H_{26}$ scintillator. The experimental data: Hertenberger et al. \cite{Her95}, Bezrukov et al. \cite{Bez73}, Boehm
et al. \cite{Boehm}, Enikeev et al. \cite{Eni87}, Kamland collaboration \cite{Kam09}, LVD experiment \cite{Agl99},
LVD-corrected \cite{Mei06} and Aglietta et al. \cite{Agl89}.} \label{Fig1}
\end{figure}

To evaluate the muon-induced neutron yield in the liquid scintillator, the
simulations described in the
Refs. \cite{Wang01,Kudry03,Arau05,Mei06} considered only the case of the muons showering within a volume of scintillator. Moreover, in them, either a mono-energetic muons beam or a real underground muons
spectrum were fired directly into a bloc of scintillator, but in all cases the
predicted neutron yield was smaller by approximately at least 30$\%$ than the measured data. Since
this observation was made with two different Monte Carlo codes- FLUKA and
Geant4- this discrepancy cannot be inferred to the models implemented in these
codes. Especially, if the cross sections of the different physics processes
in these codes agree well with the data as mentioned in the different references cited
through this article.\\
The idea investigated in this article is in order to explain the deficit noticed
in the simulations, we have to consider the contribution of the
muon-induced shower particles produced in the rock around the detector.
The cosmic muons never cross
alone over the detector. However, they are accompanied by their shower particles.
These secondary particles
will initiate different interactions, in the liquid scintillator, producing by the same
way neutrons and radioactive isotopes.
The use of shield can minimize the magnitude of the muon-induced shower particles entering
the detector, but never can suppress them. On the other hand, the shield can also play the role as a target, and then
be a source for the production of other secondary particles as described in Ref.  \cite{Arau05,Mei06}.\\
 To evaluate the muon-induced neutron yield, the simulations described in the Refs. \cite{Kudry03,Arau05} used a very large volume of
 scintillator to let the muons develop their showers, the analysis
 was made within a smaller volume of scintillator.
 For the purpose of the work presented in this article and to be close to the experimental conditions, but
at the same time to keep the generic aspect of the simulations,
a different approach is used. The size of the target scintillator was chosen to be comparable to the size of the different scintillator detectors used in the underground experiments.
 The bloc scintillator used as a target is a 10 meters cube of dodecane
 (i.e. $C_{12}H_{26}$), with density $\rho=0.75 g/cm^{3}$. The bloc
 scintillator was surrounded by 1 meter of the rock, with a standard chemical formula $SiO_{2}$ and density $\rho=2.7 g/cm^{3}$.
As mentioned in Refs. \cite{Hag00,Cri97}, for a medium as the rock, the size of 1 meter is enough to develop the muons showers.
Different beams of $10^{6}$ mono-energetic $\mu^{-}$ were
fired one at a time through the rock into the target, the energy of the muons beam was taken
within the interval of 10 to 380 GeV. A care was given to not double count the neutrons inelastically scattered.
Also, the neutrons penetrating the target after they have been created in the rock, were not counted. The
 neutrons created in the external rock constitute another aspect of the background studies, the Refs. \cite{Arau05,Mei06} give a
good perspective and a detailed study of their case.\\
The Fig.  \ref{Fig1}
shows the neutron yield obtained in the target scintillator from the Geant4 simulations and compared
to the available measurements. The agreement between the simulations and
the data is better than that is obtained in the other simulations cited above. An excellent agreement
is noticed between the actual simulations and the measurements reported by the Kamland
collaboration in the Ref. \cite{Kam09}, while the FLUKA and Geant4 simulations described in the same
reference and made in similar way as in the Refs. \cite{Wang01,Kudry03,Arau05,Mei06},
gave a neutron yield in the liquid scintillator lower by at least $ 30 \%$ than the data.
The agreement is also very good at low muon energy, with respect to the error bares.

\section{Muon-induced radioactive isotopes production}

\begin{figure}
\includegraphics[scale=0.7]{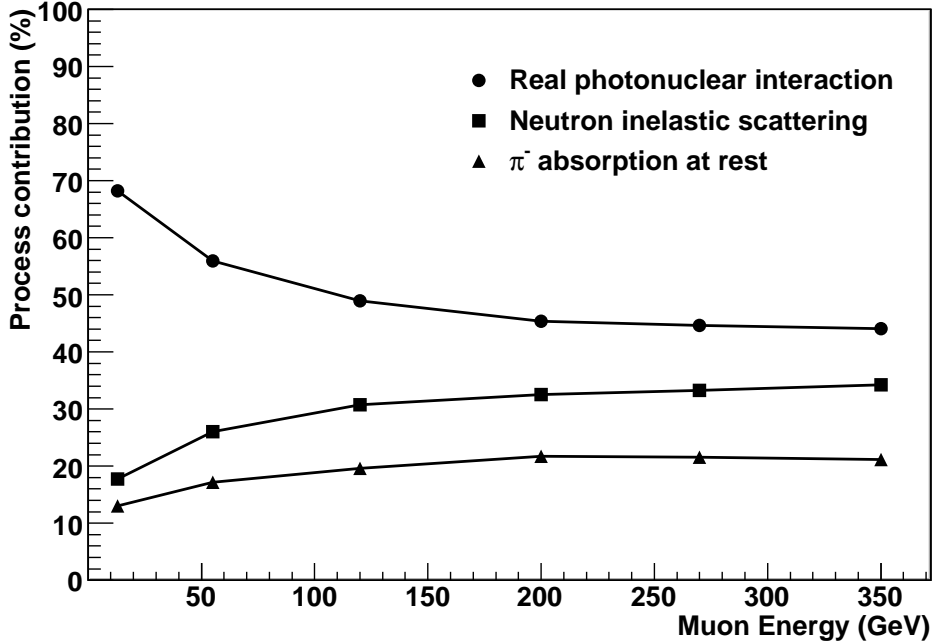}
\caption{Relative contribution of individual
 processes to the total radioactive isotope yield in $C_{12}H_{26}$
scintillator from the Geant4 simulation.} \label{Fig2}
\end{figure}

In the same study, the production of some relevant radioactive isotopes was investigated, too.
At the SPS experiment \cite{Hag00} the energy dependence of the total cross-section of the
cosmogenics radioactive isotopes was evaluated following
the power law $\sigma_{tot}(E_{\mu}) \propto E_{\mu}^{\alpha}$, where $E_{\mu}$ is the muon energy.
The tab. 1 shows their yield dependence exponent on the incident muon energy. The result of the
Geant4 is compared with the data obtained from the SPS experiment. The Fig. \ref{Fig2} shows
the main physics processes of the production of this kind of radioactive isotopes.
At low muon energy, the real photo-nuclear reaction is the dominant process. With increasing muon energy,
the contribution of the neutron inelastic scattering becomes more important, but the real photo-nuclear reaction remains
dominant. The electro-nuclear and positron-nuclear interactions, which are
not showed in the figure, contribute to less than $1\%$.

\begin{table}
\label{tab} \caption{Energy dependence exponent of the Radioactive isotopes
yield. }
\begin{tabular}{llr}
\hline
 Radioactive isotope & Energy dependence exponent \\
\cline{2-3}
& Geant4 & Data \\
\hline
$^{11}C$  & 0.6139$\pm$0.0001 &  0.70$\pm$0.16\\
$^{7}Be$ & 0.5658$\pm$0.0005  & 0.93$\pm$0.23 \\
$^{10}C$ & 0.7621$\pm$0.0004 & 0.62$\pm$0.22 \\
$^{8}Li$ & 0.5349$\pm$0.0211 &  0.50$\pm$0.71\\
$^{6}He$ & 0.6819$\pm$0.0077 &  0.71$\pm$0.22\\
$^{8}B$ & 0.3151$\pm$0.1039 & 0.84$\pm$0.45\\
\hline
\end{tabular}
\end{table}

\section{Summary}

While making the background studies of the underground experiments, the influence of the environment of the detector was checked by adding a layer of rock to a block of scintillator. Hence, for the first time with a Monte Carlo study, it was possible to confirm the experimental measurements of the muon-induced neutron yield in the liquid
scintillator.The investigation of the radioactive isotopes production shows that the dominant
process is the real photo-nuclear reaction.


\end{document}